\documentclass[conference]{IEEEtran}
\IEEEoverridecommandlockouts
\usepackage{cite}
\usepackage{algorithmic}
\usepackage{graphicx}
\usepackage{textcomp}
\usepackage{xcolor}
\def\BibTeX{{\rm B\kern-.05em{\sc i\kern-.025em b}\kern-.08em
    T\kern-.1667em\lower.7ex\hbox{E}\kern-.125emX}}

\usepackage{balance}
\usepackage{booktabs} % For formal tables
\usepackage{multirow}
\usepackage{adjustbox}
\usepackage{xcolor}
\usepackage{hyperref}
\usepackage{mfirstuc}
\usepackage{titlecaps}
\usepackage{stringstrings}
\usepackage{tabularx}
\usepackage{tabularx,booktabs} 
\usepackage{float}

\newcommand{\ignore}[1]{  }

\begin{document}

\title{Model-Based Monitoring for IoTs Smart Cities Applications\\
\thanks{This work has been partially supported by the H2020 NGPaaS project (grant n.~761557).}
}

\author{\IEEEauthorblockN{Matteo Orr\'{u}, Marco Mobilio, Anas Shatnawi, Oliviero Riganelli, Alessandro Tundo and Leonardo Mariani}
	\IEEEauthorblockA{
		%\textit{Dept. of Informatics, Systems and Communication (DISCO)} \\
		\textit{University of Milano-Bicocca}, 
		Milan, Italy \\
		\{matteo.orru, marco.mobilio, anas.shatnawi, oliviero.riganelli, leonardo.mariani\}@unimib.it}
	%\vspace{-5pt}
	a.tundo@campus.unimib.it
}

%\and
%\and
%\and
%\and
%\and

\maketitle

\begin{abstract}
	\ignore{In this work we envision a novel approach to monitoring IoTs applications, which is well suited with the Smart Cities scenario.
The main feature of our proposal is based on a quality model of the service which can frame the monitoring activity in a flexible and manageable way.} 
Smart Cities are future urban aggregations, where a multitude of heterogeneous systems and IoT devices interact to provide a safer, more efficient, and greener environment. The vision of smart cities is adapting accordingly to the evolution of software and IoT based services. The current trend is not to have a big comprehensive system, but a plethora of small, well integrated systems that interact one with each other. Monitoring these kinds of systems is challenging for a number of reasons. 
%
% For example, monitoring each sub-system separately may complicate the task of meaningful data analysis, thus reducing the possibility of optimizations, fault detection, and failure prediction. 
%
% Moreover there are different kinds of stakeholders, with different background and focus, with some of them more interested in high-level business targeted monitoring information, such as the global health and load of the system, rather than low level key performance indicators, such as free RAM memory and average CPU-Core consumption. 
%
Having a centralized and modular monitoring infrastructure, which is able to translate high level monitoring requirements into low level software metrics that must be collected, and to provide and deploy probes accordingly, can drastically ease monitoring within the context of smart cities. 
It will also help in dealing with, or at least mitigating, conflicting requirements coming from the different stakeholders involved.  
In this work, we envision a novel approach for monitoring IoT applications in a Smart City scenario, where a quality model of the services can enable the monitoring activities in a flexible and manageable way.

\end{abstract}

\vspace{5pt}


\newcommand{\anas}[1]{\textcolor{red}{{\it [Anas says: #1]}}}
\newcommand{\oliviero}[1]{\textcolor{green!20!black}{{\it [Oliviero says: #1]}}}
\newcommand{\matteo}[1]{\textcolor{blue}{{\it [Matteo says: #1]}}}
\newcommand{\marco}[1]{\textcolor{orange}{{\it [Marco says: #1]}}}
\newcommand{\leonardo}[1]{\textcolor{green!20!black}{{\it [Leo says: #1]}}}

% !TEX root =  ICities2018.tex
\section{Introduction}
\label{sec:intro}
\ignore{
\leonardo{- Se dovesse servire spazio si può ridurre la prima parte dell’introduzione che va lentamente al punto, ma se non dovesse servire possiamo tenerla così come è\\
	% - Nell’introduzione c’è qualche ref da inserire e un periodo da finire\\
	% - unirei la sezione II e III per evitare troppa frammentazione\\
	%- toglierei Figura 1 (evitando così di plagiare dal nostro paper) e metterei una bullet list indicando i 3 principali step del processo (non credo si serva altro in questo paper a parte indicare i 3 step)\\
	- lo smart city scenario potrebbe includere qualche riferimento in più al caso smart city in termini degli operatori che usano il sistema e il loro obiettivo
%	- nei ringraziamenti mettiamo solo NGPaaS visto che lo citiamo esplicitamente\\
}
}
Smart Cities represent the evolution of the current idea of urban aggregation, which leverages 
distributed, cloud systems and IoT technology to collect information to efficiently manage assets and resources. 
Among these, energy is certainly one
of the most relevant and consequently, 
%There is no doubt that a valuable resource is energy, 
there are not many doubts that energy is a relevant market for Smart Cities. 
% --- Urged by the increasingly detrimental condition of the natural enviornmental, with the ``20-20-20" directive, 
EU is aiming at lowering by 20\% both energy consumption and carbon emissions in every EU Country by 2020. 
% On the other hand, it has been estimated that 
On the other hand, traditional buildings are responsible for the 36\% of the overall carbon emissions\cite{EUbuilding}, 
which is a critical aspect to be addressed to comply with EU demands and cope with the environmental risk.
% For this reasons, in the early future, 
Consequently, the majority of EU countries will be forced to take actions to change the way they manage power consumption in buildings. 

Despite current technologies, such as SmartMetering, SmartBuilding, energy monitoring services for buildings, just to mention a few, 
can be already used to enforce policies that better address energy, security, and climate respectfulness, there is large margin for improvement. 
A case in point is represented by micro-grid technologies, which enable the management of energy within a city district (and in relation to a wider grid network) through the orchestration of a multitude of applications, sensors, actuators, meters and automation systems. 

The monitoring process of this kind of infrastructure can be challenging.  
The demand for adaptability and full integration, \emph{configurability} and \emph{programmability} cannot rule out equally relevant aspects regarding the system \emph{health}. A trade-off between flexibility and health must be addressed by taking care of both the technological and business environments. In the former
we find a multitude of heterogeneous interacting sub-systems; in the latter, a variety of business subjects and stakeholders (e.g., energy utility companies, building facilities and property managers, etc.) pursuing different goals and objectives. 
%Additionally, many different stakeholders can be involved in the process, including energy utility companies, building facilities 
%and property managers, ending up to the residential users, which makes the level of complexity increase.
%
In this work we envision a novel approach for IoT applications monitoring, which is well suited with the illustrated scenario.
We devised this approach in the context of the H2020 NGPaaS project~\cite{NGPaaS}, one of whose aims is actually to promote cloud native 
solution for 5G networks, with the aim of breaking the silo between the Telco and IT industries.
%\matteo{say something about the IoT use case?}

%which aim at achieving an high degree of cloud versatility and scalability, towards \emph{universal connectivity} of a variety of \emph{actors}, (either human or not, such as robots, devices, sensors, etc.), crossing the border that separates different realms (e.g.,. mobile, IoT, Telco, etc.) %\cite{Alleman:Rappoport:Banerjee:2010}. 
%
%This application scenario is highly demanding in terms of adaptability and full integration and \emph{configurability} and \emph{programmability} are relevant requirements. On the other hand also the system \emph{health} \textbf{must} be guaranteed. 

%This trade-off between flexibility and health must be addressed by taking care of both the technological and business environment, where, in the former, a multitude of heterogeneous interacting sub-systems evolve dynamically and, in the latter, a large variety of business subjects or stakeholders (e.g., energy utility companies, building facilities and property managers, etc.) pursuit different goals and objectives. 

The main characteristics of our proposal is that it leverages a quality model inspired by the ISO 25011 standard~\cite{ISO25011} 
to represent the concept of health for an IoT/Cloud system and relate this concept to metrics that can be measured on the actual system \cite{Shatnawi:Orru:Mobilio:Riganelli:Mariani:2018}. 
Despite the fact that solutions to collect Key Performance Indicators (KPIs) and dynamically deploy the probes are available, 
little has been done so far to connect these KPIs together into a \emph{general framework}. This would help in 
\emph{getting a better insight} on the health status of both the \emph{system} and its \emph{distinct components} 
based on \emph{operator-specific} objectives.

%Capturing the health status of a cloud system poses several challenges: (i) relating the concept of healthiness to actual KPIs in a way that is satisfactory for every actor involved in cloud systems, (ii) reporting the information about health and corresponding KPIs appropriately to cloud users, and (iii) being able to dynamically reconfigure the concept of health of the system, the KPIs that must be collected, and the deployed probes, based on dynamically emerging needs.     

Starting from the model and some knowledge of the system architecture, which is typically stored in configuration files, the definition of
monitoring goals, at various level of granularity and for different stakeholders, can be facilitated and the process of deployment of the 
probes and monitoring can be significantly automated. Moreover, since the model is aimed at capturing the goals and best-practices of a specific context, 
it can be changed at any time.
The whole process is enforced by an infrastructure called \textit{CloudHealth}, which governs each step starting from the probes' deployment, going through the 
the definition of the KPIs collected from the target system, to the visualization of the data in adaptive dashboards. 
Both the initial definition of the model and the global design of the infrastructure was provided in a previous work \cite{Shatnawi:Orru:Mobilio:Riganelli:Mariani:2018}
which can be referred for further details. In this work we outline its application in a Smart City context, where we assume the presence of a cloud and IoT infrastructure hosting IoT applications for energy management. 
%
%For instance, what the health of a system is may depend on the stakeholders and the business goals of the operators.  
%of the service which allow to perform the monitoring activities in a flexible and manageable way. 
%In this paper we present the initial results that we obtained in the design of \emph{CloudHealth}, a model-driven approach for monitoring the health of cloud systems. 
%CloudHealth uses a model inspired by the ISO 25011 standard~\cite{ISO25011} to encode the concept of health of a cloud system and to relate this concept to metrics that can be measured on the actual system. 
%
%This model is used to control the probes that must be deployed on the target system, the KPIs that are dynamically collected, and the visualization of the data in dashboards. 
%
%Interestingly, this model also represents the basis for the definition of a common language that can improve the communication between the many parties involved in the operation of a cloud system. Although we provide an initial definition of this model, the model can be changed at any time to precisely capture the goals and best-practices of specific sub-communities and sets of operators. For instance, what the health of a system is may depend on the stakeholders and the business goals of the operators.  
%
% This applies also to vendors in the IoT market, including the smart cities related applictions and hardware business.
In the following we first illustrate a realistic use case for the model-based monitoring approach (Section~\ref{sec:smart-grid-uc}). We finally describe the working principle of the propesed approach in Section~\ref{sec:model}.

%\input{background}

% !TEX root =  ICities2018.tex
\section{ A Smart City Scenario}
\label{sec:smart-grid-uc}

Energy management is one of the critical aspects of a smart city management. Micro-grids have been experienced to be able to help city districts to manage energy consumption in efficient and optimized ways. However, the way they are typically structured, in an aggregation of manifold components, either hardware (e.g., actuators, sensors, etc.) and software (e.g., standalone components, embedded software systems, mobile applications, microservices, etc.) whose interaction 
is orchestrated in a centralized fashion, makes the ordinary working scenario quite challenging when it comes to monitor the health of a system. 
Additionally, when the different stakeholders, either private or institutional, each one with their goals and perspective, come into play, this results in an further increment of the complexity at business level. 
% may come regularly into play or interact with the infrastructure. 
For example, some stakeholders may want to provision a specific kind of IoT application for their own business, which leads to different requirements in terms of monitoring, either if we consider the development stage (as in the case of DevOps) or the production one, which implies monitoring the running application and the supporting infrastructure.

\section{Model-Driven Monitoring Process} 
\label{sec:model}

%\begin{figure}[h]
%	%\begin{center}
%		\includegraphics[width=0.48\textwidth]{images/process.png}
%		\caption{The CloudHealth monitoring process.}
%		\label{fig:process1}
%	%\end{center}
%\end{figure}

The model-driven monitoring process is designed to work in the context of a Platform-as-a-Service (PaaS) cloud infrastructures that 
hosts IoT applications. % such as the above mentioned Smartmetering, Smarthome management and the like. %\cite{ferrer2016multi}, 
The PaaS infrastructure is designed to provide a high degree of cloud versatility and scalability, towards \emph{universal connectivity} 
of a variety of \emph{actors}, (either human or not, such as robots, devices, sensors, etc.), crossing the border that separates different realms 
(e.g., mobile, IoT, Telco, etc.). 

This is expected to enable the interaction between different business subjects such as IoT vendors and Vertical providers entering the IoT/cloud market. The infrastructure consists of a layered architecture where each level can be accessed by different subjects and roles. 
For instance, application developers and property managers access the top applicative/business layer (where the IoT applications are deployed and work).

%The model applies in the context of an evolved cloud Platform-as-a-Service (PaaS) infrastructures %\cite{ferrer2016multi}, 
%which aim at achieving an high degree of cloud versatility and scalability, towards \emph{universal connectivity} of a variety of \emph{actors}, (either human or not, such as robots, devices, sensors, etc.), crossing the border that separates different realms (e.g.,. mobile, IoT, Telco, etc.) %\cite{Alleman:Rappoport:Banerjee:2010}. 
%This scenario is highly demanding in terms of adaptability and full integration, \emph{configurability} and \emph{programmability} are relevant requirements. On the other hand also the system \emph{health} \textbf{must} be guaranteed. This trade-off between flexibility and health must be addressed by taking care of both the technological and business environment, where, in the former, a multitude of heterogeneous interacting sub-systems evolve dynamically and, in the latter, a large variety of business subjects or stakeholders (e.g., energy utility companies, building facilities and property managers, etc.) pursuit different goals and objectives. 

The monitoring model was created through a process of refinement and selection from the ISO/IEC 25010:2011 \cite{ISO25010} and ISO/IEC TS 25011:2017 \cite{ISO25011} standards, which provide \textit{de facto} specifications of demands widely used to evaluate the quality of generic IT services. The monitoring model represents the relations between high-level monitoring goals and the actual software metrics collected from actual services. Operators use high-level monitoring goals to easily identify the aspects to control using a dashboard. 

The mapping between goals and low-level metrics is explicitly reported in the model: this fact facilitates the automation of probes' deployment. The model specification includes information on how the high-level monitoring goals are computed from the individual metrics collected at runtime. The whole process is structured in three major steps: \textbf{(i)} Configure, \textbf{(ii)} Deploy, \textbf{(iii)} Operate.

Assuming the presence of a quality model as in our case, the first step is to \textit{select the Monitoring goals}.
This selection is a critical step and requires a certain level of knowledge of both the system and its services. 
In general, different services might differ in terms of quality demands which, in turn, are related to different KPIs. 
For example, a cloud/IoT operator may decide to monitor the reliability and the efficiency of a subset of the services available in the target platform.
This is the first stage, \textit{Configure}.
The decision made by the operator is mapped into a set of metrics specified in the monitoring model. For instance, selecting \textit{reliability} as a quality attribute, will result, according to our model, in monitoring three sub-attributes such as continuity, recoverability and availability. 

In the next stage, called \textit{Deploy}, the metrics are collected and mapped to a set of probes which are in turn deployed on the target system.
In order to do this, the infrastructure needs to have specific information on the architecture.

Finally in the last stage, \textit{Operate}, a dashboard is automatically deployed and configured in order to display the selected metrics, the associated KPIs, and the monitoring goals. This facilitates the operator who has to constantly keep an eye on the functioning of both the system and the deployed services. 
Additionally, this tool allows to perform, if necessary, a prompt diagnosis and damage control. 

Being the proposed model hierarchical, such as a tree, it describes the system, and allows different users to interact with it, at different levels of abstraction. 
A manager has an understanding of the business strategies, but usually lacks
technical knowledge. This individual would hardly understand low level KPIs, such as the \emph{data rate of successful
data delivery} over a communication link. On the other hand, he or she would rather rule out technical details about the KPIs in favour of more intelligible
monitoring goals such as ``Throughput" and ``Performance". A technician, on the contrary, would be happy to get accurate 
data, which probably represent the most relevant information for his (or her) work.

\vfill

\balance
\bibliographystyle{plain} % ACM-Reference-Format}
\bibliography{soheal2018} 

\end{document}